\documentclass[aps,prl,twocolumn,superscriptaddress]{revtex4-1}
\usepackage{graphicx}

\begin{document}

\title{Noise induced stabilization in population dynamics}

\author{Matthew Parker}
\affiliation{School of Physics and Astronomy, University of Minnesota, Minneapolis, MN 55455}

\author{Alex Kamenev}
\affiliation{School of Physics and Astronomy, University of Minnesota, Minneapolis, MN 55455}
\affiliation{William I. Fine Theoretical Physics Institute, University of Minnesota,
Minneapolis, MN 55455}

\author{Baruch Meerson}
\affiliation{Racah Institute of Physics, Hebrew University of Jerusalem, Jerusalem 91904, Israel}

\begin{abstract}
We investigate a model where strong noise in a sub-population creates a metastable state in an otherwise unstable two-population system. The induced metastable state is vortex-like, 
and its persistence time grows exponentially with the noise strength. 
A variety of distinct scaling relations are observed depending on the relative strength of the sub-population noises.
\end{abstract}

\maketitle


The phenomenon of noise induced metastability \cite{gitterman,spagnolo,*spagnolo2,mielke} is of importance in  ecology \cite{guttal} and plant biology \cite{d2005} and has found practical applications in engineering \cite{ibrahim}.
The typical models  \cite{gitterman,spagnolo2,mielke} consider periodically modulated
one-dimensional (1d) stochastic systems.
The modulation renders the system to be deterministically unstable during a part of the modulation period. An external noise can prevent the escape for several successive periods of external modulation, trapping the system into a metastable state.  As a result, the noise causes an increase of the system persistence time by a factor compared to the noiseless case.

Here we consider a different model with two stochastic degrees of freedom, which we call $x$ and $y$.  The $y$ degree of freedom (e.g. imbalance between the numbers of two competing gene alleles) undergoes a strong and fast noise which conserves the total population size $x$. The latter experiences a slow evolution under the
influence of a deterministic potential $V(x)$ along with a sign-definite feedback from the population size imbalance $\propto y^2$ and a relatively weak (demographic) noise. We show that, even if the $x$-dynamics itself is unstable and prone to a rapid escape, the strong $y$-noise can lock it in an {\em exponentially} long-lived vortex-like metastable state. The corresponding exponent exhibits a variety of non-trivial scaling regimes, depending on the relative strength of the noises in the $x$ and $y$ subsystems. A similar model was shown to describe a two-patch Lotka-Volterra system \cite{abta}. More distantly related models were recently
discussed in the context of biochemical regulatory networks \cite{assaf2} and nanomechanical oscillators \cite{Dykman}.

Our model can be cast into the universal form
\begin{eqnarray}\label{x-y}
 &&   \dot x = -V^{\prime}(x)-y^2 +\xi_x(t)\,, \nonumber \\
 &&    \dot y =-2y + \xi_y(t)\, ; \\
 &&  \langle \xi_{x(y)}(t)\xi_{x(y)}(t')\rangle = 2T_{x(y)}\delta(t-t') \,, \nonumber
\end{eqnarray}
where $V^{\prime}=dV(x)/dx$, and $T_x$ and $T_y$ characterize the noise strength in the total and differential population size, respectively. The interesting regime of parameters is  $T_x< T_y$. The noise effects are substantial when the population is close to a bifurcation point.
In this case the properly rescaled deterministic potential takes the form
\begin{equation}\label{force}
    V(x)= -x^3/3-\delta x,
\end{equation}
where $\delta$ is the bifurcation parameter, and we have shifted the $x$ variable to have the bifurcation point at $x=0$.

\begin{figure}
\begin{center}
\includegraphics[width=2.9in]{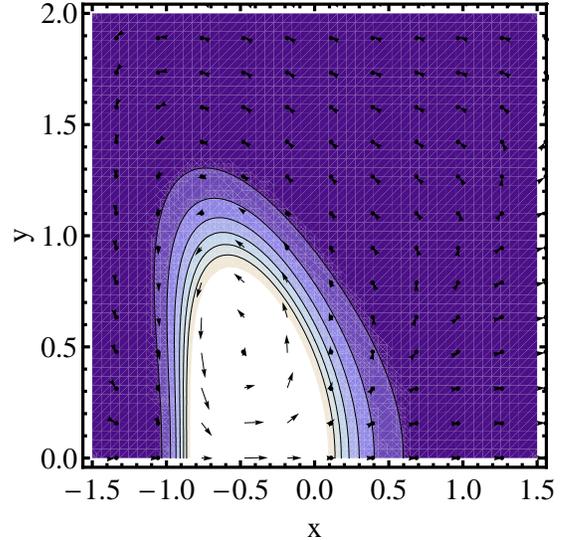}
\end{center}
\caption[]{\label{fig:mf_stream} (Color online) The quasi-stationary FP state. The contours represent the probability density $P(x,y)$; the arrows show the probability current density. The system is symmetric around $y=0$, and only the top half is shown.  $T_x=0.05, T_y=0.5$. }
\end{figure}

A simple realization of this model is provided by two species  $A$ and $B$, which undergo the reaction $A + B \stackrel{\lambda}{\rightarrow} 2X$, where $X$ is either $A$ or $B$. This is the well-known Moran  process for modeling neutral genetic drift \cite{moran}. This is the fastest process which conserves the total population size.
In addition, the population size may slowly evolve according to e.g. the following set of reactions $X \stackrel{\beta_\mp}{\leftrightarrow} 0$ and
$A + B \stackrel{\alpha}{\rightarrow} A + B + X$.  In this case $x=(n_A+n_B-N)/N$ and $y=(n_A-n_B)/N$, where $N=\beta_-/\alpha$ is the population
size close to the bifurcation, and $\delta = 4\alpha\beta_+/\beta_-^2-1$ is the bifurcation parameter. In the limit of large population size $N\gg 1$, the corresponding Master equation can be approximated by a Fokker-Planck equation~\cite{vanKampen}. In the vicinity of the bifurcation point, the latter reads
\begin{eqnarray}\label{eqn:FP}
\dot{P}(x,y) = \!\!&-&\!\!\!\partial_x\! \left[ \left(-V^{\prime}(x) - y^2 \right) P(x,y) - \frac{2}{N} \partial_x P(x,y) \right] \nonumber \\
&-&\!\!\partial_y \left[ -2 y P(x,y) - \left( \lambda + \frac{2}{N} \right) \partial_y P(x,y) \right],
\end{eqnarray}
where $P(x,y,t)$ is the probability distribution function, and time is measured in units of $2/\beta_-$.
Equation~(\ref{eqn:FP}) is equivalent to the Langevin equations (\ref{x-y}), where the two ``temperatures'' are given by $T_x=2/N$ and $T_y=T_x+2\lambda/\beta_-$. When the drift rate $\lambda$ is fast, one has the strong inequality $T_x\ll T_y$.

First we focus on the case of exact bifurcation,  $\delta=0$. The $x$-equation takes the form
$\dot x =x^2-y^2+\xi_x$. Without noise the $y$-variable tends
to zero, leading to $\dot x=x^2$ dynamics in the $x$-direction. This has $x=0$ as the marginally stable point. An arbitrarily weak $x$-noise is sufficient to kick the system out of this fixed point and set it on the path to unlimited proliferation, $x\to \infty$. One may think
thus that the $\delta=0$ system is destined to blow up in a very short time. Recall, however, that the $y$-noise is substantial. Although $\langle y\rangle=0$, the mean square value $\langle y^2\rangle>0$ and is large compared with $T_x$. One can then expect the $x$-dynamics to be governed by the effective potential $V_\mathrm{eff}(x)=-x^3/3+ \langle y^2\rangle x$. This potential exhibits a minimum at $x=-\sqrt {\langle y^2\rangle}$, and a maximum at $x=\sqrt{ \langle y^2\rangle}$. As a result, a long-lived metastable distribution, peaked at $x=-\sqrt {\langle y^2\rangle}$, can be created.
A numerical solution of the Fokker-Planck (FP) equation~(\ref{eqn:FP}) supports this expectation. Figure~\ref{fig:mf_stream} shows the slowly varying quasi-stationary distribution observed at late times. Notably, the probability currents develop two counter-rotating vortices.  Before reaching the point $x=y=0$, the ``particle" is kicked in the $y$-direction, where the $x$-evolution is directed toward
population contraction. The vortices, therefore, arrest the population explosion. We note that probability current vortices in non-equilibrium \emph{stationary} states -- the Brownian vortices --
were recently observed in experiment \cite{sun}.

\begin{figure}
\begin{center}
\includegraphics[width=3.2in]{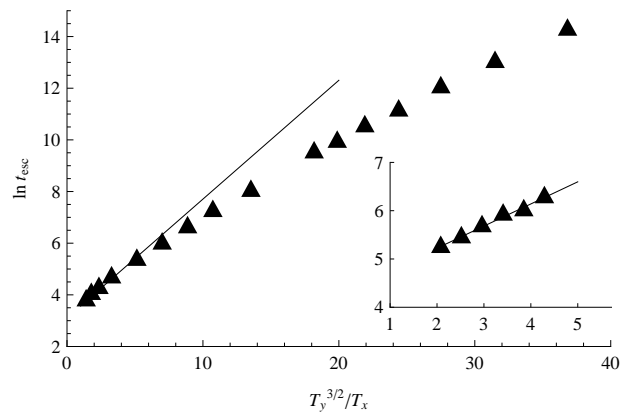}
\end{center}
\caption[]{\label{fig:act_data_skew} Simulated escape times for $T_y = 0.05$ and varying $T_x$. The straight line has slope of $\sqrt{2}/3$, cf. Eq.~(\ref{t-escape}). Inset: the limit $\sqrt{T_x} > T_y$.}
\end{figure}

Our main goal is to evaluate the lifetime of such a noise-induced vortex-like metastable state. We start from  qualitative considerations. As a first approximation one can estimate the mean-square $y$-deviation in the harmonic potential $y^2$, cf. Eq.~(\ref{x-y}), as
$\langle y^2\rangle =T_y/2$. Therefore $V_\mathrm{eff}^\mathrm{max} - V_\mathrm{eff}^\mathrm{min}= \sqrt{2}\,T_y^{3/2}/3$, and  one expects that
\begin{equation}\label{t-escape}
    \ln t_\mathrm{esc}\simeq \sqrt{2}\,T_y^{3/2}/3T_x\,.
\end{equation}
Remarkably, the escape time is exponentially {\em increasing} with the $y$-noise strength $T_y$, while exhibiting the standard Arrhenius scaling with $T_x$. Our numerical simulations of the Langevin Eqs.~(\ref{x-y}), see Fig.~\ref{fig:act_data_skew},  confirm Eq.~(\ref{t-escape}) as long as $T_y^{3/2}/T_x$ is not too large. At larger values of this parameter, however, Eq.~(\ref{t-escape}) greatly overestimates the lifetime of the metastable state.

The reason for this deviation is that for large $T_y$ the typical potential barrier is too high for the $x$-motion to overcome. Then,  instead of
relying  on typical realizations of $y$-noise, the system  prefers to wait for a rare $y$-trajectory which stays anomalously close to $y=0$.
The probability that the $y$-motion is confined to the interval $|y(t)|<y_0$ for a time $t_0$ is given by $\exp[-E_y(y_0)t_0]$. Here $E_y$ is the lowest eigenvalue of the 1d 
FP equation in the $y$-direction with absorbing boundary conditions at $y=\pm y_0$. $E_y$ can be estimated as $E_y \propto T_y/y_0^2$, where $T_y$ is the proper diffusion coefficient. On the other hand, the probability that during the time interval $t_0$ the $x$-coordinate will diffuse from $x=-y_0$ to $x=+y_0$ is given by $\exp(-y_0^2/T_x t_0)$, where $T_x$ is the proper diffusion coefficient. Maximizing the product of these two probabilities with respect to $y_0^2/t_0$, one finds  that the probability of the optimal rare fluctuation scales as $\exp(-\sqrt{T_y/T_x})$. These estimates suggest that $\ln t_\mathrm{esc}\propto \sqrt{T_y/T_x}$ once $\sqrt{T_y/T_x}<T_y^{3/2}/T_x$, i.e.
$\sqrt{T_x}<T_y$. This behavior is indeed qualitatively consistent with  Fig.~\ref{fig:act_data_skew}.

To put these considerations on a more quantitative basis we shall assume that the dynamics can be separated into the fast
$y$-motion and slow $x$-motion. The latter adiabatically adjusts to the instantaneous value of $y^2(t)$.  We then solve an auxiliary
problem of finding the probability of $y$-trajectories with a given functional form $\langle y^2\rangle =y_0^2(t)$. Here  $y_0(t)$ is an arbitrary
{\em slow} function of time, such that $y_0(\pm \infty)= \sqrt{T_y/2}$, while the averaging is taken over the fast $y$-fluctuations. Integrating over an intermediate  time-scale $\Delta t$ that is long relative to these fluctuations one can then write $\int_{t-\Delta t}^{t+\Delta t}[ y_0^2(t) - y^2(t)] dt = 0$.
We can thus introduce the functional constraint $\delta\big(\int[ y^2(t) -y_0^2(t)] dt\big)$ into the stochastic functional integral over ${\cal D}y$ \cite{MSR,DeDominicis,Janssen} and elevate it into the exponent with the help of the auxiliary {\em slow} field $\chi(t)$. As a result we obtain an
effective Lagrangian
\begin{equation}\label{lagrangian-y}
    {\cal L}_y = \frac{(\dot y +2y)^2}{4T_y} - \chi (y^2-y_0^2)\,,
\end{equation}
where the $\chi$-integration runs from $-i\infty$ to $i\infty$. Employing the slowness of the $\chi(t)$ field, the Gaussian integral over the fast $y(t)$
can be evaluated using the Fourier transformation.  This leads to an effective Lagrangian for $\chi$ in the form
 \begin{equation}\label{lagrangian-y-2}
    {\cal L}_\chi\! =\chi y_0^2-\!\! \int\!\frac{d\omega}{2\pi}\, \ln\! \left(\!1\!+\frac{4T_y\chi}{\omega^2+4}\right)\!  =
    \chi y_0^2 + 1-\!\sqrt{T_y\chi+1}\!.
\end{equation}
Finally, the $\chi$-integration can be evaluated in the saddle point  approximation: $\chi(t)=T_y/4y_0^4-T_y^{-1}$. This yields the probability
of $y$-motion conditioned on $\langle y^2\rangle =y_0^2(t)$:
\begin{equation}\label{P-y}
    P[y_0]\propto e^{-\int\! dt\, E_y(y_0)},\quad\quad E_y(y_0) = \frac{T_y}{4y_0^2}-1+\frac{y_0^2}{T_y} \,.
\end{equation}
Notice that $E_y$ is non-negative and equal to zero if and only if $y_0^2=T_y/2$. Therefore, the condition
$y_0^2(\pm \infty)=T_y/2$ is necessary for convergence of the integral in Eq.~(\ref{P-y}). The saddle point calculation is justified
as long as $\int\! dt\, E_y[y_0(t)]\gg 1$.

Having found the conditional probability of $y$-motion with a given profile of $ \langle y^2\rangle$, we turn now to the
$x$-degree of freedom. According to the scale separation assumption, it is governed by the Langevin equation
$\dot x =x^2-y_0^2(t)+\xi_x(t) $, where $y_0^2(t)$ is a slow function of time with $y_0^2(\pm \infty)=T_y/2$ and $y_0^2(0)<T_y/2$.
Our goal is to evaluate the escape rate of the $x$-variable from its metastable minimum at $x=-\sqrt{T_y/2}$ during the time when $y_0^2(t)$ is
suppressed with respect to its asymptotic values. We then maximize this escape rate, taken with weight $P[y_0]$, Eq.~(\ref{P-y}), against the optimal
time-dependent variance $y_0(t)$.

\begin{figure}
\begin{center}
\includegraphics[width=3in]{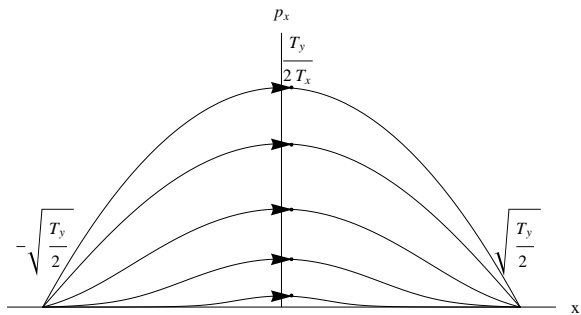}
\end{center}
\caption[]{\label{fig:escape_paths} Phase  portrait of the optimal escape paths for different $T_x$. Lower lines correspond to lower values of $T_x$. The momentum has been rescaled by $T_y/T_x$ so that all paths would coincide if they followed simple activation [assumed by Eq.~(\ref{t-escape})].}
\end{figure}

\begin{figure}
\begin{center}
\includegraphics[width=3.2in]{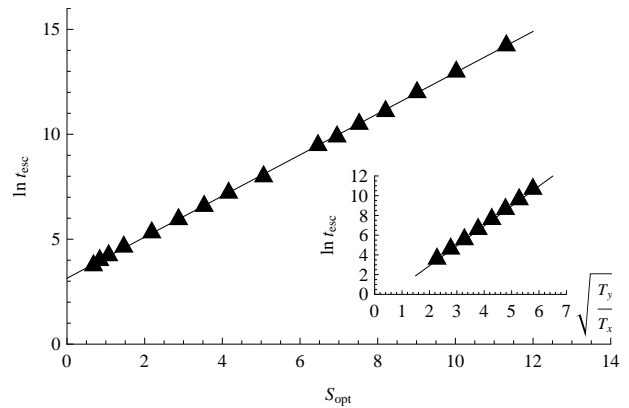}
\end{center}
\caption[]{\label{fig:log_tau_v_action} Simulated escape times vs. the optimal action for $T_y = 0.05$ and varying $T_x$. Inset: the limit $(T_y^2/T_x)^{1/6} \gg 1$ (with $T_y = 1000)$. The straight line is Eq.~(\ref{t-escape-1}).}
\end{figure}

Since the escape rate in the $x$-direction  is expected to be small, it can be found through a semiclassical treatment of the corresponding FP equation
\cite{freidlin,*dykman1984,*graham}. The  proper FP Hamiltonian has the form
\begin{equation}\label{hamiltonian}
    {\cal H}[x,p_x;y_0(t)]=p_x[-V^{\prime}(x) - y_0^2+T_xp_x]-E_y[y_0(t)]\,,
\end{equation}
where $x$ and $p_x$ are canonically conjugate variables, and $y_0(t)$ is an external time-dependent parameter. The last term accounts for the
statistical weight of a realization of $y_0(t)$, given by $P[y_0]$, Eq.~(\ref{P-y}). If $y_0(t)$ is an adiabatically slow
function, the escape proceeds along the zero-energy trajectory of this Hamiltonian, which connects the two fixed points $(-\sqrt{T_y/2}, 0)$ and
$(+\sqrt{T_y/2}, 0)$ on its $(x,p_x)$ phase plane, Fig.~\ref{fig:escape_paths}. Putting $V(x)=-x^3/3$, one finds for the (slowly varying in time) optimal trajectory
\begin{equation}\label{eqn:momentum}
p_x(x;y_0) = \frac{1}{2 T_x}\left[ y_0^2  - x^2 +
\sqrt{( y^2_0  - x^2)^2 + 4 T_x E_y(y_0)} \right].
\end{equation}
The corresponding escape time, within exponential accuracy, is given by the classical action, i.e. the area of the phase plane
under the zero-energy trajectory
\begin{equation}\label{eqn:action}
\ln t_\mathrm{esc}[y_0]= S[y_0] = \int_{-\sqrt{T_y/2}}^{\sqrt{T_y/2}} p_x(x;y_0)\, dx .
\end{equation}
The final step is to find the optimal $y_0$ realization. This is achieved by demanding $\delta S[y_0]/\delta y_0=0$,
solving for an implicit function of time $y_0 =y_0(x)$ and substituting it back into Eq.~(\ref{eqn:action}). This leads to the optimal action, $S_\mathrm{opt}$, and corresponding
escape time $\ln t_\mathrm{esc}= S_\mathrm{opt}$. In Fig.~\ref{fig:log_tau_v_action} this escape time is compared with our Monte-Carlo simulations results, and an excellent agreement is observed.

It is easy to show that, for $T_y\ll \sqrt{T_x}$, the optimal $y_0$ tends to $\sqrt{T_y/2}$ and thus $E_y\to 0$, while
$S_\mathrm{opt}=\sqrt{2}\,T^{3/2}_y/3T_x$. We thus recover Eq.~(\ref{t-escape}). In the opposite limit $T_y\gg \sqrt{T_x}$, one finds
$y_0(0)\ll \sqrt{T_y/2}$. One can thus simplify Eq.~(\ref{P-y}) as $E_y\approx T_y/4y_0^2$. With this substitution Eq.~(\ref{eqn:momentum}) can be
simplified by the rescaling $x=\tilde x(T_xT_y)^{1/6}$ and    $y_0=\tilde y_0(T_xT_y)^{1/6}$, which brings the action (\ref{eqn:action}) into the form $S=\sqrt{T_y/T_x}\int \tilde p(\tilde x; \tilde y_0)d\tilde x$. The integration limits are $\pm (T_y^2/T_x)^{1/6}\to \pm \infty$ in the limit of interest, while $\tilde p=\left[\tilde{y}_0^2  - \tilde{x}^2 +  \sqrt{(\tilde{y}_0^2  - \tilde{x}^2)^2 + \tilde{y}_0^{-2}}\, \right]/2$ is a parameterless  function. Optimizing it over $\tilde y_0$ and performing $\tilde x$-integration,  one finds
\begin{equation}\label{t-escape-1}
    \ln t_\mathrm{esc}= \frac{2 \pi}{3} \sqrt{T_y\over T_x}\,, \quad\quad \quad     \sqrt{T_x}\ll T_y\,,
\end{equation}
which confirms our qualitative estimates below Eq.~(\ref{t-escape}) and provides the numerical factor. The latter is compared with our
Monte-Carlo simulations in the inset of Fig.~\ref{fig:log_tau_v_action}. Notice that the actual condition for the applicability of the asymptotic result (\ref{t-escape-1}) is $1\ll (T_y^2/T_x)^{1/6}$. Again, the population lifetime {\em increases} with the  $y$-noise strength. Notice also that the normal Arrhenius scaling in the parameter $T_x$ gives way to a stretched exponential law with ${T_x}^{-1/2}$. A similar transition is  known  as Efros-Shklovskii law~\cite{Shklovskii} in the context of hopping transport in disordered semiconductors.

We consider now deviations from the exact bifurcation point, i.e. $\delta \neq 0$. If $|\delta| \gg T_y$, the deterministic ``force", cf. Eq.~(\ref{force}),  is very strong, and the $y$-noise is not important for the system's persistence time. At $\delta = \delta_c = T_y/2$ the effective force associated with the $y$-noise is canceled by the deterministic $\delta$-force. This causes a noise-induced shift in the bifurcation of the $x$-dynamics. That is, it is much harder to destabilize the population in the presence of strong $y$ noise.
In the vicinity of  this noise-shifted bifurcation one finds the standard scaling of the lifetime $\ln t_\mathrm{esc}= 4 \,(\delta_c - \delta )^{3/2}/ 3 T_x $, cf. Eq~(\ref{t-escape}).

\begin{figure}
\begin{center}
\includegraphics[width=3.2in]{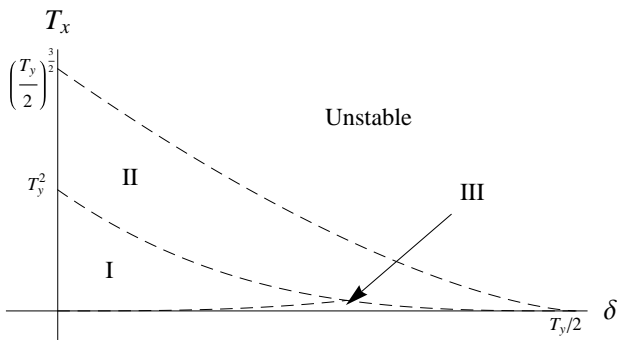}
\end{center}
\caption[]{\label{fig:phase_diagram} Phase diagram of the system as a function of $T_x$ and $\delta$ with $T_y$ held constant. The dashed lines represent crossovers between different scaling relations.  In region I,  $\ln t_\mathrm{esc}\! = 2 \pi / 3 \sqrt{T_y / T_x}$. In region II,  $\ln t_\mathrm{esc}\! = 4  \left(T_y / 2 - \delta \right)^{3/2} / 3 T_x$. In region III,  $\ln t_\mathrm{esc}\! = \pi T_y / 2 \delta^{3/2}$.}
\end{figure}

On the other hand, away from the the noise-shifted bifurcation, i.e. at $|\delta -\delta_c|/\delta_c > (\sqrt{T_x}/T_y)^{2/3}$, the scaling changes qualitatively. To find the new scaling we look for the zero energy trajectory of the Hamiltonian~(\ref{hamiltonian}) with $\delta\neq 0$ and $E_y=T_y/4y_0^2$ and optimize the action over $y_0(x)$, as explained above. In this way we find
\begin{equation}\label{t-escape-delta}
    \ln t_\mathrm{esc}\!= \frac{2 \pi}{3} \sqrt{\frac{T_y}{T_x}} \ \ \mathcal{S}\left[\frac{\delta}{(T_x T_y)^{1/3}}\right] \,, \quad    \sqrt{T_x}\ll T_y\,,
\end{equation}
where the universal function $\mathcal{S}(\tilde \delta)$ has the following asymptotic limits: $\mathcal{S}(\tilde\delta) \approx 1 + 0.71 \tilde\delta$ for $\tilde\delta \ll 1$ and $\mathcal{S}(\tilde\delta) \approx   3 \ \tilde\delta^{-3/2}/4$ for $\tilde\delta\gg 1$.  This means that the
scaling of the population lifetime given by Eq.~(\ref{t-escape-1}) is basically intact as long as $\delta \lesssim (T_x T_y)^{1/3}$. In the opposite limit, the escape time scales as
\begin{equation}\label{eqn:hidscaling}
 \ln t_\mathrm{esc} = \pi T_y / 2 \delta^{3/2} .
\end{equation}
This is independent of $T_x$. For $\delta > (T_x T_y)^{1/3}$ the system can escape  even at $T_x = 0$ via paths with unusually small $y$. Figure~\ref{fig:phase_diagram} shows the regions where the three scaling relations~(\ref{t-escape}), (\ref{t-escape-1}), (\ref{eqn:hidscaling}) are valid.

In summary, we have studied a novel system where strong noise creates metastability. Increasing the noise strength $T_y$ increases the  lifetime of the (vortex-like) metastable state. Escape from this state is governed by a variety of scaling relations depending on the relative role of $T_x$ (the strength of noise in the total population size) and $T_y$ (the strength of noise in the differential size). 

We are grateful to J. Krug and M. Dykman for valuable discussions.
This research was supported by NSF Grant DMR-0804266 and  U.S.-Israel Binational Science Foundation Grant 2008075.

\bibliography{xy-pap}

\begin{thebibliography}{20}%
\makeatletter
\providecommand \@ifxundefined [1]{%
 \@ifx{#1\undefined}
}%
\providecommand \@ifnum [1]{%
 \ifnum #1\expandafter \@firstoftwo
 \else \expandafter \@secondoftwo
 \fi
}%
\providecommand \@ifx [1]{%
 \ifx #1\expandafter \@firstoftwo
 \else \expandafter \@secondoftwo
 \fi
}%
\providecommand \natexlab [1]{#1}%
\providecommand \enquote  [1]{``#1''}%
\providecommand \bibnamefont  [1]{#1}%
\providecommand \bibfnamefont [1]{#1}%
\providecommand \citenamefont [1]{#1}%
\providecommand \href@noop [0]{\@secondoftwo}%
\providecommand \href [0]{\begingroup \@sanitize@url \@href}%
\providecommand \@href[1]{\@@startlink{#1}\@@href}%
\providecommand \@@href[1]{\endgroup#1\@@endlink}%
\providecommand \@sanitize@url [0]{\catcode `\\12\catcode `\$12\catcode
  `\&12\catcode `\#12\catcode `\^12\catcode `\_12\catcode `\%12\relax}%
\providecommand \@@startlink[1]{}%
\providecommand \@@endlink[0]{}%
\providecommand \url  [0]{\begingroup\@sanitize@url \@url }%
\providecommand \@url [1]{\endgroup\@href {#1}{\urlprefix }}%
\providecommand \urlprefix  [0]{URL }%
\providecommand \Eprint [0]{\href }%
\@ifxundefined \urlstyle {%
  \providecommand \doi  [0]{\begingroup \@sanitize@url \@doi}%
  \providecommand \@doi [1]{\endgroup \@@startlink {\doibase
  #1}doi:\discretionary {}{}{}#1\@@endlink }%
}{%
  \providecommand \doi  [0]{doi:\discretionary{}{}{}\begingroup
  \urlstyle{rm}\Url }%
}%
\providecommand \doibase [0]{http://dx.doi.org/}%
\providecommand \Doi [0]{\begingroup \@sanitize@url \@Doi }%
\providecommand \@Doi  [1]{\endgroup\@@startlink{\doibase#1}\@@Doi}%
\providecommand \@@Doi [1]{#1\@@endlink}%
\providecommand \selectlanguage [0]{\@gobble}%
\providecommand \bibinfo  [0]{\@secondoftwo}%
\providecommand \bibfield  [0]{\@secondoftwo}%
\providecommand \translation [1]{[#1]}%
\providecommand \BibitemOpen [0]{}%
\providecommand \bibitemStop [0]{}%
\providecommand \bibitemNoStop [0]{.\EOS\space}%
\providecommand \EOS [0]{\spacefactor3000\relax}%
\providecommand \BibitemShut  [1]{\csname bibitem#1\endcsname}%
\bibitem [{\citenamefont {Dayan}\ \emph {et~al.}(1992)\citenamefont {Dayan},
  \citenamefont {Gitterman},\ and\ \citenamefont {Weiss}}]{gitterman}%
  \BibitemOpen
  \bibfield  {author} {\bibinfo {author} {\bibfnamefont {I.}~\bibnamefont
  {Dayan}}, \bibinfo {author} {\bibfnamefont {M.}~\bibnamefont {Gitterman}}, \
  and\ \bibinfo {author} {\bibfnamefont {G.}~\bibnamefont {Weiss}},\
  }\href@noop {} {\bibfield  {journal} {\bibinfo  {journal} {Phys. Rev. A},\
  }\textbf {\bibinfo {volume} {46}},\ \bibinfo {pages} {757} (\bibinfo {year}
  {1992})}\BibitemShut {NoStop}%
\bibitem [{\citenamefont {Mantegna}\ and\ \citenamefont
  {Spagnolo}(1996)}]{spagnolo}%
  \BibitemOpen
  \bibfield  {author} {\bibinfo {author} {\bibfnamefont {R.}~\bibnamefont
  {Mantegna}}\ and\ \bibinfo {author} {\bibfnamefont {B.}~\bibnamefont
  {Spagnolo}},\ }\href@noop {} {\bibfield  {journal} {\bibinfo  {journal}
  {Phys. Rev. Lett.},\ }\textbf {\bibinfo {volume} {76}},\ \bibinfo {pages}
  {563} (\bibinfo {year} {1996})}\BibitemShut {NoStop}%
\bibitem [{\citenamefont {Fiasconaro}\ and\ \citenamefont
  {Spagnolo}(2009)}]{spagnolo2}%
  \BibitemOpen
  \bibfield  {author} {\bibinfo {author} {\bibfnamefont {A.}~\bibnamefont
  {Fiasconaro}}\ and\ \bibinfo {author} {\bibfnamefont {B.}~\bibnamefont
  {Spagnolo}},\ }\href@noop {} {\bibfield  {journal} {\bibinfo  {journal}
  {Phys. Rev. E},\ }\textbf {\bibinfo {volume} {80}},\ \bibinfo {pages}
  {041110} (\bibinfo {year} {2009})}\BibitemShut {NoStop}%
\bibitem [{\citenamefont {Mielke}(2000)}]{mielke}%
  \BibitemOpen
  \bibfield  {author} {\bibinfo {author} {\bibfnamefont {A.}~\bibnamefont
  {Mielke}},\ }\href@noop {} {\bibfield  {journal} {\bibinfo  {journal} {Phys.
  Rev. Lett.},\ }\textbf {\bibinfo {volume} {84}},\ \bibinfo {pages} {818}
  (\bibinfo {year} {2000})}\BibitemShut {NoStop}%
\bibitem [{\citenamefont {Guttal}\ and\ \citenamefont
  {Jayaprakash}(2007)}]{guttal}%
  \BibitemOpen
  \bibfield  {author} {\bibinfo {author} {\bibfnamefont {V.}~\bibnamefont
  {Guttal}}\ and\ \bibinfo {author} {\bibfnamefont {C.}~\bibnamefont
  {Jayaprakash}},\ }\href@noop {} {\bibfield  {journal} {\bibinfo  {journal}
  {Ecol. Modell.},\ }\textbf {\bibinfo {volume} {201}},\ \bibinfo {pages} {420}
  (\bibinfo {year} {2007})}\BibitemShut {NoStop}%
\bibitem [{\citenamefont {D'Odorico}\ \emph {et~al.}(2005)\citenamefont
  {D'Odorico}, \citenamefont {Laio},\ and\ \citenamefont {Ridolfi}}]{d2005}%
  \BibitemOpen
  \bibfield  {author} {\bibinfo {author} {\bibfnamefont {P.}~\bibnamefont
  {D'Odorico}}, \bibinfo {author} {\bibfnamefont {F.}~\bibnamefont {Laio}}, \
  and\ \bibinfo {author} {\bibfnamefont {L.}~\bibnamefont {Ridolfi}},\
  }\href@noop {} {\bibfield  {journal} {\bibinfo  {journal} {Proc. Natl. Acad.
  Sci. USA},\ }\textbf {\bibinfo {volume} {102}},\ \bibinfo {pages} {10819}
  (\bibinfo {year} {2005})}\BibitemShut {NoStop}%
\bibitem [{\citenamefont {Ibrahim}(2006)}]{ibrahim}%
  \BibitemOpen
  \bibfield  {author} {\bibinfo {author} {\bibfnamefont {R.}~\bibnamefont
  {Ibrahim}},\ }\href@noop {} {\bibfield  {journal} {\bibinfo  {journal} {J.
  Vib. Control},\ }\textbf {\bibinfo {volume} {12}},\ \bibinfo {pages} {1093}
  (\bibinfo {year} {2006})}\BibitemShut {NoStop}%
\bibitem [{\citenamefont {Abta}\ \emph {et~al.}(2007)\citenamefont {Abta},
  \citenamefont {Schiffer},\ and\ \citenamefont {Shnerb}}]{abta}%
  \BibitemOpen
  \bibfield  {author} {\bibinfo {author} {\bibfnamefont {R.}~\bibnamefont
  {Abta}}, \bibinfo {author} {\bibfnamefont {M.}~\bibnamefont {Schiffer}}, \
  and\ \bibinfo {author} {\bibfnamefont {N.}~\bibnamefont {Shnerb}},\
  }\href@noop {} {\bibfield  {journal} {\bibinfo  {journal} {Phys. Rev.
  Lett.},\ }\textbf {\bibinfo {volume} {98}},\ \bibinfo {pages} {98104}
  (\bibinfo {year} {2007})}\BibitemShut {NoStop}%
\bibitem [{\citenamefont {Assaf}\ and\ \citenamefont {Meerson}(2008)}]{assaf2}%
  \BibitemOpen
  \bibfield  {author} {\bibinfo {author} {\bibfnamefont {M.}~\bibnamefont
  {Assaf}}\ and\ \bibinfo {author} {\bibfnamefont {B.}~\bibnamefont
  {Meerson}},\ }\href@noop {} {\bibfield  {journal} {\bibinfo  {journal} {Phys.
  Rev. Lett.},\ }\textbf {\bibinfo {volume} {100}},\ \bibinfo {pages} {58105}
  (\bibinfo {year} {2008})}\BibitemShut {NoStop}%
\bibitem [{\citenamefont {Atalaya}\ \emph {et~al.}()\citenamefont {Atalaya},
  \citenamefont {Isacsson},\ and\ \citenamefont {Dykman}}]{Dykman}%
  \BibitemOpen
  \bibfield  {author} {\bibinfo {author} {\bibfnamefont {J.}~\bibnamefont
  {Atalaya}}, \bibinfo {author} {\bibfnamefont {A.}~\bibnamefont {Isacsson}}, \
  and\ \bibinfo {author} {\bibfnamefont {M.}~\bibnamefont {Dykman}},\
  }\href@noop {} {\bibinfo  {journal} {arXiv:1103.2758}}\BibitemShut {NoStop}%
\bibitem [{\citenamefont {Moran}(1962)}]{moran}%
  \BibitemOpen
\bibfield  {journal} {  }\bibfield  {author} {\bibinfo {author} {\bibfnamefont
  {P.}~\bibnamefont {Moran}},\ }\href@noop {} {\emph {\bibinfo {title} {{The
  Statistical Processes of Evolutionary Theory}}}}\ (\bibinfo  {publisher}
  {Clarendon Press, Oxford},\ \bibinfo {year} {1962})\BibitemShut {NoStop}%
\bibitem [{\citenamefont {van Kampen}(1992)}]{vanKampen}%
  \BibitemOpen
  \bibfield  {author} {\bibinfo {author} {\bibfnamefont {N.}~\bibnamefont {van
  Kampen}},\ }\href@noop {} {\emph {\bibinfo {title} {{Stochastic Processes in
  Physics and Chemistry}}}}\ (\bibinfo  {publisher} {North Holland,
  Amsterdam},\ \bibinfo {year} {1992})\BibitemShut {NoStop}%
\bibitem [{\citenamefont {Sun}\ \emph {et~al.}(2009)\citenamefont {Sun},
  \citenamefont {Lin}, \citenamefont {Darby}, \citenamefont {Grosberg},\ and\
  \citenamefont {Grier}}]{sun}%
  \BibitemOpen
  \bibfield  {author} {\bibinfo {author} {\bibfnamefont {B.}~\bibnamefont
  {Sun}}, \bibinfo {author} {\bibfnamefont {J.}~\bibnamefont {Lin}}, \bibinfo
  {author} {\bibfnamefont {E.}~\bibnamefont {Darby}}, \bibinfo {author}
  {\bibfnamefont {A.}~\bibnamefont {Grosberg}}, \ and\ \bibinfo {author}
  {\bibfnamefont {D.}~\bibnamefont {Grier}},\ }\href@noop {} {\bibfield
  {journal} {\bibinfo  {journal} {Phys. Rev. E},\ }\textbf {\bibinfo {volume}
  {80}},\ \bibinfo {pages} {010401(R)} (\bibinfo {year} {2009})}\BibitemShut
  {NoStop}%
\bibitem [{\citenamefont {Martin}\ \emph {et~al.}(1973)\citenamefont {Martin},
  \citenamefont {Siggia},\ and\ \citenamefont {Rose}}]{MSR}%
  \BibitemOpen
  \bibfield  {author} {\bibinfo {author} {\bibfnamefont {P.}~\bibnamefont
  {Martin}}, \bibinfo {author} {\bibfnamefont {E.}~\bibnamefont {Siggia}}, \
  and\ \bibinfo {author} {\bibfnamefont {H.}~\bibnamefont {Rose}},\ }\href@noop
  {} {\bibfield  {journal} {\bibinfo  {journal} {Phys. Rev. A},\ }\textbf
  {\bibinfo {volume} {8}},\ \bibinfo {pages} {423} (\bibinfo {year}
  {1973})}\BibitemShut {NoStop}%
\bibitem [{\citenamefont {de~Dominicis}(1976)}]{DeDominicis}%
  \BibitemOpen
  \bibfield  {author} {\bibinfo {author} {\bibfnamefont {C.}~\bibnamefont
  {de~Dominicis}},\ }\href@noop {} {\bibfield  {journal} {\bibinfo  {journal}
  {J. Physique (Paris)},\ }\textbf {\bibinfo {volume} {37}},\ \bibinfo {pages}
  {1} (\bibinfo {year} {1976})}\BibitemShut {NoStop}%
\bibitem [{\citenamefont {Janssen}(1976)}]{Janssen}%
  \BibitemOpen
  \bibfield  {author} {\bibinfo {author} {\bibfnamefont {H.}~\bibnamefont
  {Janssen}},\ }\href@noop {} {\bibfield  {journal} {\bibinfo  {journal} {Z.
  Physik B.},\ }\textbf {\bibinfo {volume} {23}},\ \bibinfo {pages} {377}
  (\bibinfo {year} {1976})}\BibitemShut {NoStop}%
\bibitem [{\citenamefont {Freidlin}\ and\ \citenamefont
  {Wentzell}(1984)}]{freidlin}%
  \BibitemOpen
  \bibfield  {author} {\bibinfo {author} {\bibfnamefont {M.}~\bibnamefont
  {Freidlin}}\ and\ \bibinfo {author} {\bibfnamefont {A.}~\bibnamefont
  {Wentzell}},\ }\href@noop {} {\emph {\bibinfo {title} {{Random Perturbations
  of Dynamical Systems}}}}\ (\bibinfo  {publisher} {Springer-Verlag, Berlin},\
  \bibinfo {year} {1984})\BibitemShut {NoStop}%
\bibitem [{\citenamefont {Dykman}\ and\ \citenamefont
  {Krivoglaz}(1984)}]{dykman1984}%
  \BibitemOpen
  \bibfield  {author} {\bibinfo {author} {\bibfnamefont {M.}~\bibnamefont
  {Dykman}}\ and\ \bibinfo {author} {\bibfnamefont {M.}~\bibnamefont
  {Krivoglaz}},\ }\href@noop {} {\bibfield  {journal} {\bibinfo  {journal}
  {{"}Physics Reviews{"}},\ }\textbf {\bibinfo {volume} {5}},\ \bibinfo {pages}
  {266} (\bibinfo {year} {1984})}\BibitemShut {NoStop}%
\bibitem [{\citenamefont {Graham}(1989)}]{graham}%
  \BibitemOpen
  \bibfield  {author} {\bibinfo {author} {\bibfnamefont {R.}~\bibnamefont
  {Graham}},\ }\href@noop {} {\emph {\bibinfo {title} {in {"}Theory of
  Continuous Fokker--Planck Systems{"}}}},\ edited by\ \bibinfo {editor}
  {\bibfnamefont {F.}~\bibnamefont {Moss}}\ and\ \bibinfo {editor}
  {\bibfnamefont {P.}~\bibnamefont {McClintock}}\ (\bibinfo  {publisher}
  {Cambridge University Press, Cambridge},\ \bibinfo {year} {1989})\
  Chap.~\bibinfo {chapter} {7}, p.\ \bibinfo {pages} {225}\BibitemShut
  {NoStop}%
\bibitem [{\citenamefont {Shklovskii}\ and\ \citenamefont
  {Efros}(1984)}]{Shklovskii}%
  \BibitemOpen
  \bibfield  {author} {\bibinfo {author} {\bibfnamefont {B.}~\bibnamefont
  {Shklovskii}}\ and\ \bibinfo {author} {\bibfnamefont {A.}~\bibnamefont
  {Efros}},\ }\href@noop {} {\emph {\bibinfo {title} {{Electronic Properties of
  Doped Semiconductors}}}},\ Vol.~\bibinfo {volume} {1}\ (\bibinfo  {publisher}
  {Springer-Verlag, Berlin},\ \bibinfo {year} {1984})\BibitemShut {NoStop}%
\end{thebibliography}%

\end{document}